# The Crucial Role of Ground- and Space-Based Remote Sensing Studies of Cometary Volatiles in the Next Decade (2023-2032)

Nathan X. Roth[1,2], NASA GSFC, 301.286.8151, nathaniel.x.roth@nasa.gov

Dennis Bodewits[3], Boncho Bonev[4], Anita Cochran[5], Michael Combi[6], Martin Cordiner[1], Neil Dello Russo[7], Michael DiSanti[1], Sara Faggi[1,4], Lori Feaga[8], Yan Fernandez[9], Manuela Lippi[1,4], Adam McKay[1,4], Matthew Knight[10], Stefanie Milam[1], John W. Noonan[11], Anthony Remijan[12], Geronimo Villanueva[1]

[1]NASA Goddard Space Flight Center, Greenbelt, MD, USA
[2]Universities Space Research Association, Columbia, MD, USA
[3]Physics Department, Leach Science Center, Auburn University, Auburn, AL, USA
[4]American University, Washington DC, USA
[5]McDonald Observatory, The University of Texas at Austin, Austin, TX, USA
[6]University of Michigan, Ann Arbor, MI, USA
[7]Johns Hopkins University Applied Physics Laboratory, Laurel, MD, USA
[8]University of Maryland, College Park, MD, USA
[9]University of Central Florida, Orlando, FL, USA
[10]United States Naval Academy, Annapolis, MD, USA
[11]Lunar and Planetary Laboratory, University of Arizona, Tucson, AZ, USA
[12]National Radio Astronomy Observatory, Charlottesville, VA, USA

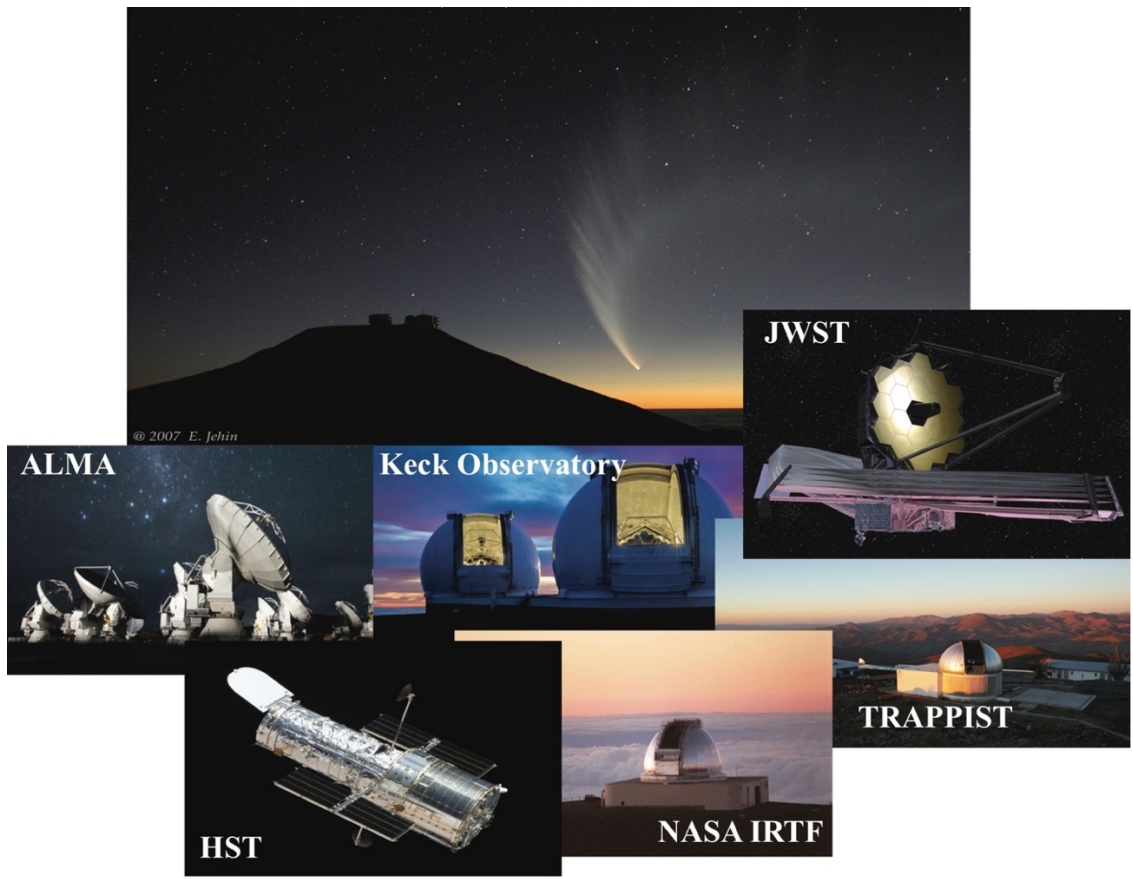

# The Crucial Role of Remote Sensing Studies of Cometary Volatiles in the Next Decade

**Abstract:** The study of comets affords a unique window into the birth, infancy, and subsequent history of the solar system. There is strong evidence that comets incorporated pristine interstellar material as well as processed nebular matter into their nuclei, providing insights into the composition and prevailing conditions over wide swaths of the solar nebula at the time of planet formation. Their populations bear a record spanning the life of the solar system. Dynamically new Oort cloud comets harbor primitive ices that have been stored thousands of astronomical units from the Sun and have suffered minimal thermal or radiative processing since their emplacement ~4.5 Gyr ago. Periodic, more dynamically evolved comets such as the Halley-type and Jupiter-family comets reveal the effects of lives spent over a range of heliocentric distances, including perihelion passages into the very inner solar system. Systematically characterizing the information imprinted in the native ice compositions of these objects is critical to understanding the formation and evolution of the solar system, the presence of organic matter and water on the terrestrial planets, the chemistry present in protoplanetary disks around other stars, and the nature of interstellar interlopers such as 2I/Borisov. Although comet rendezvous and sample return missions can provide remarkable insights into the properties of a few short-period comets, the on-sky capacity necessary to perform population-level comet studies while simultaneously remaining sensitive to the paradigm-challenging science that individual comets can reveal can only be provided by remote sensing observations. Here we report the state-of-the-art in ground- and space-based remote sensing of cometary volatiles, review the remarkable progress of the previous decade, articulate the pressing questions that ground- and space-based work will address over the next ten years, and advocate for the technology and resources necessary to realize these aspirations.

## 1. Review of the Previous Decade in Remote Sensing Cometary Science

Cometary nuclei contain some of the most primitive solar system material available for routine study. Their composition is predominantly inferred through remote sensing of coma gases using ground- and space-based facilities. From the microwave to x-rays, each wavelength regime probes a different physical and chemical domain of the coma. Here we review the remarkable progress in the study of comets at each of these wavelengths in the previous decade.

*Optical Studies.* The compositions of comets have been studied with optical spectra for over 150 years. Optical wavelengths sample spectral features from fragment species (e.g., CN, OH, $NH_2$, $CO^+$)[1] produced in the coma through photolysis or other processing. The bright emission features of these species facilitate study of the largest sample size among the different wavelength regimes. The development of comet-specific narrowband filters and low-resolution spectroscopy enabled large observing surveys and the development of comprehensive taxonomies[2]. More recently, the robotic TRAPPIST telescopes have enabled long-term and continuous monitoring of comets through dual hemispheric sky coverage and comet-specific narrowband filters. In the past decade, optical observations have been obtained with high spectral resolution instruments. These spectra allow study of individual lines of molecular bands that yield important information on the chemical processes that control the sublimation of the cometary nucleus ices. The large subset of optical studies has also enabled numerous studies of isotopic ratios such as $^{15}N/^{14}N$.

*Ultraviolet studies.* Light with wavelengths shorter than 300 nm is effectively blocked by the Earth's atmosphere and requires an orbital observatory. UV wavelengths contain the emission from electronic transitions of atoms and molecules, providing a unique window on the regime where molecules are excited, dissociate, and react with other particles. UV observations are critical to understand inner coma emission physics as well as to measure the general atomic and molecular abundances of the coma. These important diagnostics include access to fragments that are more easily detectable than their progenitor species (H and OH for $H_2O$; $CO_2^+$; multiple sulfur-bearing fragments)[3]; emission from reactions that are highly diagnostic of electron impact dissociation rates and local plasma temperatures; fragment species that can be used as proxies for species that do not have a dipole moment (atomic oxygen for $O_2$); and in the Extreme UV, the charge exchange interaction of solar wind alpha particles with the neutral gas in small body atmospheres.

The SWAN instrument onboard NASA's Solar and Heliospheric Observatory (SOHO) provides a key survey of cometary activity, targeting Ly-α emission from the hydrogen envelope surrounding comets on large angular scales[3]. Spectroscopy with a range of orbital facilities, including the Hubble Space Telescope and Far Ultraviolet Spectroscopic Explorer, has been used to measure the $CO/H_2O$ production rates of comets and interstellar objects, to determine their $CO_2$ content, to search for noble gases, and to investigate physical processes in the coma[4]. These measurements are crucial for differentiating the cometary properties that are primordial, those that are the result of alteration, and those that are an effect of emission mechanisms in the coma.

*Mm/Sub-mm Studies.* From the ground, the microwave/sub-mm region provides access to the largest subset of known cometary molecules through their pure rotational transitions[5]. The in-situ Rosetta MIRO sub-mm instrument mapped the near-nucleus coma of comet 67P in unprecedented detail[6], but ground-based mm-wave observations continue to provide a more systematic method for measuring detailed compositions. A unique advantage of the mm/sub-mm range is the sensitivity to emission from organic molecules with a greater degree of complexity than detectable in other wavelength regions, including astrobiologically relevant species such as formamide and cyanoacetylene. Access to isotopic ratios including D/H and $^{15}N/^{14}N$ is another major benefit, crucial in tracing the thermochemical history of primitive solar system matter.

The advent of ALMA in 2012 sparked a new era of high-resolution interferometric studies of the coma, enabling mapping molecules released directly from the nucleus, as well as identifying the in-situ synthesis of gas-phase species as a result of coma chemistry[7]. The wideband receiver at IRAM has enabled molecular surveys of many comets, resulting in the first detections of cometary ethyl alcohol and glycoaldehyde[8], and 18 cm OH observations at facilities such as Nançay have provided water production rates in over one hundred comets[9]. Heterodyne receiver technology enables the derivation of high-resolution line-of-sight velocity information, from which 3D coma structures can be determined, allowing the physical and chemical properties of comets to be robustly constrained. Asymmetric outgassing patterns provide insights into cometary activity, $CH_3OH$ rotational temperature mapping reveals detailed thermal mechanisms at play, and mapping daughter species helps reveal the macromolecular material in the nucleus[10,11].

*Near-IR Studies.* High resolution ($\lambda/\Delta\lambda > 2 \times 10^4$), long-slit spectrographs operating in the 1-5 μm region permit systematic measurement of at least nine trace volatiles together with $H_2O$, including uniquely the symmetric hydrocarbons $C_2H_2$, $CH_4$ and $C_2H_6$ that lack allowed rotational transitions. The relative strengths of emission lines and their variations over a cross-section of coma subtended by a slit provides information about the relative abundances of these simple molecules and how they are released from the nucleus. Observations over the last twenty-five years have led to the emergence of a continually evolving compositional taxonomy[12,13,14], establishing unambiguous variability in abundance ratios of native ices among the approximately 40 comets measured to date, as well as relationships between the ices stored in comets.

Recent significant advances in the sensitivity and spectral coverage of ground-based IR spectrographs have enabled the cometary database to expand quickly, increasing the number of comets observed and the frequency of observations. In late 2016, the cross-dispersed iSHELL facility spectrograph at the NASA Infrared Telescope Facility replaced its predecessor (CSHELL).

Combined with IRTF's daytime observing capability (unique among ground-based IR platforms) this significantly expanded such compositional studies to additional (weaker) comets, including those with limited solar elongation angles and heliocentric distances well within 1 au[15,16,17].

## 2. Pressing Questions for the Next Decade

With its multi-wavelength coverage, large on-sky availability, and increasing capacity for long-term studies, remote sensing will be fundamental in addressing pressing questions in cometary science in the upcoming decade. These include taking detailed inventory of coma composition including trace gases to better constrain formation conditions, assessing taxonomies across all wavelengths to aid in understanding photochemical links between parent and daughter species, interpreting the temporal and spatial variability of coma composition to identify the nature of volatile release, and discerning the link between comets, asteroids, and interstellar objects.

*Inventory of coma composition and constraints on formation conditions.* A key missing component in ground-based studies of native ices is $CO_2$, likely one of the major volatiles present in comets, but which is not observable owing to atmospheric opacity. A spectroscopic survey using the Akari Infrared Satellite detected $CO_2$ in 17 of 18 comets observed[18], establishing its prominence in comets. This was reinforced by the flyby of 103P/Hartley 2 during the Deep Impact extended mission (EPOXI), which revealed $CO_2$ as a localized driver of activity that dragged icy grains into the coma[19], and by time-resolved measurements of $CO_2$ by Rosetta at 67P[20,21]. **The James Webb Space Telescope (JWST) will afford an unprecedented view into the $CO_2$ content of comets alongside CO and $H_2O$, the three primary drivers of cometary activity.**

Comets provide a unique link to the chemistry of the protoplanetary disk from which the solar system formed. Comets represent the best class of objects for studying this link because they have retained the richest inventory of volatiles (ices) accreted in the cold outer regions of the disk. Understanding comets as a population is complementary to studies of disks in extrasolar planetary systems in that they represent midplane material that is not easily accessed by modern observatories, further emphasizing the role of comets as remnants preserving material from this midplane region. **Thus, we advocate for surveys of cometary composition to help constrain protoplanetary disk ice-phase chemistry and the natal reservoirs for comet formation.**

*Combining Independent Taxonomies and Understanding Coma Photochemistry.* In the past few decades, comet compositional surveys have expanded from UV through radio, resulting in taxonomies of comets at most wavelengths[2,5,13]. All comets have similar spectra, but careful examination shows that each has unique relative volatile abundances. However, surveys at different wavelengths cannot yet be fully inter-compared. More than 250 comets have been observed in the optical, yet IR and radio surveys each have less than 50 comets sampled, primarily owing to more recent development of increasingly sensitive instruments. Thus, there is a limited overlap between the comets included in optical taxonomies and those in the IR or the radio.

**During the next decade, we advocate giving high priority to developing a consistent taxonomy for comets across all wavelengths.** This will require coordinating multi-wavelength campaigns on the most suitable targets. Observing fewer comets but at multiple wavelengths adds more information to our understanding than observing more comets in a single spectral regime; it allows us a more complete understanding of the relationship between nucleus ices and coma products. These multi-wavelength campaigns will require support from observatories to efficiently schedule. Such a comprehensive taxonomy will allow us to map in detail the storage and release of cometary volatiles and connect it to compositional variations within the solar nebula.

# The Crucial Role of Remote Sensing Studies of Cometary Volatiles in the Next Decade

Owing to the lack of overlap in optical studies (where many products have strong transitions) with those at radio or IR wavelengths (where putative parents can be sampled), the lineages of many coma molecules (e.g., CN, $C_2$, $C_3$) are not definitively understood. The photochemistry of larger molecules such as OCS, HNC, and $H_2CO$, which show evidence of production via unknown coma sources[22], is poorly constrained. **We advocate for increased organization of simultaneous observations of daughter molecules with their prospective parents through multi-wavelength campaigns to improve our knowledge of these photochemical relationships.**

*Temporal Variability of Coma Composition.* In early observations of the gas comae of comets, data were often obtained in single, "snap-shot" observations, characterizing the observed composition at only a specific instant in time and at a single heliocentric distance. Additionally, all spacecraft missions to comets before Rosetta were "snap-shots", as they consisted of brief flybys. However, improvements in instrumentation have enabled observing comets over larger portions of their orbits and larger ranges of heliocentric distances. For short period comets, similar observations obtained during different apparitions have become more common[16,23]. Long term observations revealed that cometary coma composition is not static, but can vary over the course of an apparition, or from apparition to apparition. The Rosetta mission provided unprecedented insight into variable coma composition at 67P/Churyumov-Gerasimenko, detailing variability on the scale of hours to weeks and months, linked to diurnal, seasonal, and episodic behavior[24,25]. Examples abound from ground-based observations as well[26,27,28]. **These orbital trends observed for past comets illustrate the future need to observe comets throughout their orbits to better understand cometary composition and the evolution and variability of cometary activity.**

Volatiles stored in the nucleus are not necessarily released into the coma by direct sublimation. The EPOXI mission at 103P/Hartley 2, along with complementary ground-based observations, demonstrated that sublimation of $H_2O$ from icy coma grains contributed significantly to overall gas production[19,29]. In several cases, such as for 73P/Schwassmann-Wachmann 3B, ground-based observations indicated that icy grain sublimation dominated $H_2O$ production[30], and there is direct and indirect evidence of icy coma grains in many other comets[31]. Understanding these phenomena in comets and contextualizing the results of missions such as EPOXI is critical to properly interpreting the nature of the coma, and can only be accomplished through future remote sensing.

Additionally, much work remains in understanding how trace species are associated or segregated in nucleus ices. The EPOXI and Rosetta missions both revealed comets in which different species originated from distinct locations on the nucleus[20,21,32,33]. Analogous ground-based results can be derived from interferometric maps[7] or long-slit near-IR spectroscopy[28]. **Determining population-level trends in the associations of volatiles in comets over the next decade will shed light on the nature of volatile release and improve our understanding of the question *"How are comets put together?"*[34].**

*Understanding the Link Between Comets and Asteroids and Context for Interstellar Objects.* The discovery of "main-belt comets" (MBCs), or asteroids displaying activity best explained by sublimation, has challenged the notion that asteroids and comets share separate formation histories and ignited a search to characterize the composition and nature of their outgassing. No volatiles have been detected in MBCs to date[35], likely a consequence of the very low activity expected for MBCs and the sensitivity of current facilities. **Coupled with its freedom from telluric interference, the exceptional sensitivity of JWST will offer the best chance at detecting $H_2O$ or other parent volatiles released from these enigmatic objects and begin to unravel the source(s) of their activity[35,36] and their relationship to comets.**

# The Crucial Role of Remote Sensing Studies of Cometary Volatiles in the Next Decade

The visits of interstellar objects 1I/'Oumuamua and 2I/Borisov provided an excellent example of the role that ground-based and orbiting observations will play for this newly discovered class of objects. Although no coma was conclusively detected for 1I/'Oumuamua[37], 2I/Borisov presented a clear gas and dust coma and was available for observations at perihelion and beyond. Optical wavelength campaigns provided the first confirmation of outgassing in 2I/Borisov with the detection of CN[38], followed by [OI][39] and OH[40] (tracers for $H_2O$), and provided molecular and dust production rates as far as 100 days after perihelion[41]. The first parent volatiles in an interstellar comet were detected shortly thereafter, with sub-mm lines of HCN and CO detected with ALMA[42] and CO through the Fourth Positive system with HST[43]. **With rendezvous missions to these little-understood objects extremely challenging, we advocate for remote sensing studies which will be critical to characterizing those interstellar objects passing through the solar system in the next decade and decoding the clues they carry to planet formation around other stars.**

## 3. Requisite Technology, Studies, and Resources

Realizing that these aspirations will require efforts on multiple fronts, we advocate addressing the technological and theoretical challenges necessary to support the next decade of cometary science and urge institutional support for the increasing complexity of these studies.

*Theoretical Challenges.* The next decade in cometary science will afford the opportunity to more directly link the gases characterized by remote sensing with native nucleus ices. Enabling these advances will require theoretical and laboratory work, including detailed modeling of subsurface outgassing, collision rates, excitation rates, and cross-sections, as well as increasingly sophisticated studies of coma physics. **We advocate for significant attention to address the theoretical and laboratory work necessary to more fully interpret remote sensing observations, including those linking parent and fragment species.**

*Smaller Facility Access Loss.* Significant challenges for access to smaller facilities have arisen, including increased privatization and loss of institutional support. 10-m (and larger) class facilities lack the flexibility for the comprehensive and serial observations of comets necessary to address many pressing questions in the next decade. Smaller facilities at all wavelengths, such as the NASA-IRTF, the Arizona Radio Observatory, and small (< 10-m) optical telescopes will be crucial to providing this temporal coverage and responding to newly discovered comets. **We strongly advocate for increased support of these smaller facilities in the coming decade.**

*Increased Access to 10-m Class and Astrophysical Facilities.* Solar system programs constitute a small portion of time on 10-m class/astrophysical facilities, yet they can provide breakthrough science along with public outreach[44]. For example, the recently upgraded NIRSPEC-2 on Keck II was used to observe past and likely future mission target 46P/Wirtanen during its historic close approach to Earth in 2018[45]. **We strongly advocate supporting these large astrophysical facilities for general comet studies, to complement as well as to contextualize both past and future missions.** The use of Keck is critical for studies of space mission targets as well as for testing key mission findings over many comets, as these cannot all be visited by missions. We recommend that comet science at Keck is prioritized for measurements that demand (1) the highest possible sensitivity in combination with the longest slits currently available in the IR (at Keck 2) for in-depth spatial studies, and/or (2) truly simultaneous observations of product species with their hypothesized (yet highly uncertain) precursor molecules. This is possible only with coordinated multiwavelength studies, such as those using Keck or ESO-VLT.

*Programs to Address Archival Data.* Significant archival data for comets have accumulated from facilities at all wavelengths. From unaddressed data to observations that would benefit from

# The Crucial Role of Remote Sensing Studies of Cometary Volatiles in the Next Decade

secondary analysis using upgraded algorithms, the opportunities for archival studies to contribute to our understanding of comets over the next decade is substantial[14]. However, the community currently lacks a program that supports comprehensively addressing these data. **We advocate for increased support for and focus on the analysis of archival cometary data.**

*Comets and Big Data.* Increasingly complex facilities (e.g., ALMA) and new instruments offering long on-source integrations and serial observations of comets (ZTF, Rubin Observatory) are leading cometary science into the realm of "big data". ALMA comet observations can exceed several terabytes for a single object, and long-term studies of comets with large echellograms (e.g., iSHELL) produce complex data sets. Leveraging current and future advanced facilities will require addressing the enormous data that they generate, including support for computational resources, data reduction algorithm development, and the analysis of large observing programs and surveys. **We strongly advocate for an avenue to secure these resources in the coming decade.**

## 4. Future Facilities to Enable Cometary Studies

Breakthrough science for comets with current state-of-the-art facilities has already been described in detail. Next generation observatories are being built, designed, and/or conceived for unprecedented sensitivity at multiple wavelengths and allowing for new discoveries in comets.

*Ground-Based Facilities.* Substantial breakthroughs in understanding coma chemistry will be achieved in the next decade with improved sensitivities (and larger bandwidths) at radio wavelengths, leading to the detection and mapping of new species that will constrain the chemical history of our solar system, as well as providing insight into the chemical reagents available for life in primitive planetary environments. **To realize these goals, we recommend investment in mm/sub-mm ground-based facilities,** with focus on larger, more sensitive single-dish facilities, as exemplified by the Atacama Large-Aperture Submillimeter Telescope (AtLAST), presently undergoing design in Europe, and on the expansion of existing interferometers (ALMA/VLA).

The upcoming extremely large (20–40m) ground-based facilities such as the European-Extremely Large Telescope (E-ELT), the Giant Magellan Telescope (GMT), and the Thirty Meter Telescope (TMT) will provide unparalleled sensitivity and resolution for cometary science at optical and near-IR wavelengths. Instruments such as HIRES and METIS at E-ELT, NIRES at TMT, and GMTNIRS at GMT will offer excellent wavelength coverage for cometary volatile studies. **We strongly advocate supporting cometary science at these next-generation facilities, including rapid-follow up capability and comet-specific (e.g., CN) narrowband filters.**

*Space-Based Facilities.* The James Webb Space Telescope (JWST) is the next great observatory currently slated to launch in 2021. This 6.5m diameter telescope will be the largest infrared observatory ever operated in space, covering wavelengths from 0.6-28.5 μm with low to medium spectral resolution, offering sensitivity that exceeds similar observatories by orders of magnitude[46]. The extreme sensitivity coupled to the wavelength coverage will enable revolutionary science of faint comets inaccessible to ground-based facilities, detailed studies of comet activity beyond the water sublimation point, and discoveries yet to be determined.

Studies for future space telescopes are underway for the 2030's. Proposed concepts include a large UV-optical-NIR observatory (LUVOIR) and the far infrared observatory (Origins Space Telescope). These observatories will enable new science for comets through significant enhancement in collection area and advanced instrumentation. For example, LUVOIR will be able to image comet nuclei before their activity obscures them and determine abundances of water and other species with high precision. The Origins telescope will have access to water rotational and ro-vibrational lines to determine the abundances and isotopic ratios of volatiles[47.]

# The Crucial Role of Remote Sensing Studies of Cometary Volatiles in the Next Decade

The observations made possible with these new space observatories will surpass the performance of previous missions such as Hubble, Spitzer, and Herschel and we strongly advocate that provisions to support cometary science with these platforms should be considered throughout the mission planning and implementation stages.

## 5. Conclusion

The upcoming decade will afford opportunities to reveal the composition, nature, and history of comets like never before. Advances in technology will enable studies of "ordinary" (moderately bright) comets that are currently only feasible for blockbusters such as Hale-Bopp. The temporal and wavelength coverage of ground-based facilities, coupled with the sensitivity of future space-based facilities, will place the exceptional phenomena observed in several comets into the context of the broader population, as well as reveal unexpected and paradigm-challenging new science. Unraveling the nature of comets (and in turn, the birth and evolution of our solar system) will require continued support and buy-in from the astronomical community.

*Relevant White Papers.*
Solar System Science with JWST – Hammel et al.
Planetary Science with Astrophysics Assets – Bauer et al.
Cryogenic Cometary Sample Return – Westphall et al.
Main Belt Comets as Clues to the Distribution of Water in the Early Solar System – Meech et al.
Understanding Solar System Formation Through Small Body Exploration – Davidsson et al.
Volatile Sample Return in the Solar System – Milam et al.
Isotopic Ratios in Water and the Origin of Earth's Oceans – Lis et al.
The Case for Non-Cryogenic Comet Nucleus Sample Return – Keiko Nakamura-Messenger et al.
Solar System Science with Space Telescopes – Juanola-Parramon et al.